\documentclass[aps,pra,twocolumn,superscriptaddress,longbibliography]{revtex4-1}
\usepackage{graphicx}
\usepackage{natbib}
\usepackage{amsmath}

\newcommand{\TJWat}{IBM T.J. Watson Research Center, Yorktown Heights, NY 10598, USA}
\newcommand{\IBMZ}{IBM Research - Zurich, 8803 Rueschlikon, Switzerland}

\newcommand{\ket}[1]{\ensuremath{\left|{#1}\right\rangle}}

\newcommand{\figfolder}[1]{}

\usepackage[usenames,dvipsnames]{xcolor}
\usepackage[dvips,letterpaper=true,pagebackref=false]{hyperref}
\hypersetup{
	pdfnewwindow=true, 
	colorlinks=true, 
	linkcolor=MidnightBlue,
	citecolor=Magenta,
	urlcolor=RoyalBlue
}

\begin{document}

\title{A universal gate for fixed-frequency qubits via a tunable bus}

\author{David C. McKay}
\email{dcmckay@us.ibm.com}
\affiliation{\TJWat}
\author{Stefan Filipp}
\affiliation{\IBMZ}
\author{Antonio Mezzacapo}
\affiliation{\TJWat}
\author{Easwar Magesan}
\affiliation{\TJWat}
\author{Jerry M. Chow}
\affiliation{\TJWat}
\author{Jay M. Gambetta}
\affiliation{\TJWat}

\date{\today}

\begin{abstract}
	A challenge for constructing large circuits of superconducting qubits is to balance addressability, coherence and coupling strength. High coherence can be attained by building circuits from fixed-frequency qubits, however, leading techniques cannot couple qubits that are far detuned. Here we introduce a method based on a tunable bus which allows for the coupling of two fixed-frequency qubits even at large detunings. By parametrically oscillating the bus at the qubit-qubit detuning we enable a resonant exchange (XX+YY) interaction. We use this interaction to implement a 183~ns two-qubit iSWAP gate between qubits separated in frequency by $854~\rm{MHz}$ with a measured average fidelity of 0.9823(4) from interleaved randomized benchmarking. This gate may be an enabling technology for surface code circuits and for analog quantum simulation. 
\end{abstract}

\maketitle

\section{Introduction}

Superconducting qubits are a promising implementation for fault-tolerant quantum computing~\cite{gambetta:2015}, however, proposed circuits will be large --- a logical qubit in the surface code could require up to $10^4$ physical qubits~\cite{fowler:2012}. Building these large circuits requires highly coherent and strongly interacting physical qubits to achieve high-fidelity gates. At the same time, unwanted interactions which undermine the fault-tolerance built into the surface code must be minimized. These divergent conditions on coherence, interaction and crosstalk have led to two main qubit architectures depending on which condition is given highest priority.

In the first approach the qubit frequencies are tunable and interactions are controlled by dynamically tuning pairs of qubits into and out of specific resonance conditions~\cite{dicarlo:2009,bialczak:2010}. Although this enables fast gates with relatively high on/off ratios, these qubits are susceptible to dephasing noise from the tunability channel, typically flux noise, which lowers coherence~\cite{koch:2007}. Furthermore, this approach is sensitive to frequency crowding; as a pair of qubits tune into resonance they must avoid crossing through resonances with other qubits. Utilizing longitudinal interactions (see e.g., Ref.~\cite{kerman:2008,wang:2009,kerman:2013,billangeon:2015,didier:2015,richer:2016}) may alleviate these crowding issues, but interactions of this type have yet to be implemented.

The second approach is to use fixed-frequency qubits, which have demonstrated superior coherence properties in circuits implemented using two~\cite{sheldon:2015b} and three-dimensional~\cite{paik:2011,rigetti:2012} architectures. A number of gates have been proposed and realized for fixed-frequency qubits by applying one or more microwave drives~\cite{blais:2007,leek:2009,poletto:2012,chow:2013b,rigetti:2005}. In particular, the cross-resonance (CR) gate~\cite{rigetti:2010,chow:2011} has demonstrated fidelities greater than 0.99~\cite{sheldon:2015b}. However, similar to many drive-activated gates, it is only effective when the qubits are closely spaced compared to the anharmonicity (the detuning between the qubit transition and the transition to the next excited state). For the transmon qubit, used here and in the plurality of experiments, this limits the frequency spacing to approximately a few hundred MHz. For large circuits, this is a challenging constraint for fabrication, crosstalk and addressability. \\ 

Ideally we would like to combine the best aspects from both approaches: the flexibility and scalability of tunable qubits with the coherence and fidelity of fixed-frequency qubits. This is possible by transferring tunability from the computational qubits to the coupling degree of freedom thereby reducing sensitivity to noise. There are two implementations of a tunable coupler, direct and indirect. A direct tunable coupler is realized by a tunable circuit element between qubits, e.g., a flux-tunable inductor~\cite{bertet:2006,ploeg:2007,chen:2014,wulschner:2015,allman:2014}. Alternatively, an indirect tunable coupling is realized by fixed coupling to a tunable resonator. When the qubits are far detuned from the resonator, i.e. in the dispersive limit of the circuit quantum electrodynamics architecture, this arrangement realizes a tunable bus and the exchange coupling between the qubits can be tuned by changing the qubit-bus detuning~\cite{niskanen:2006,wallquist:2006,harrabi:2009,wang:2011,whittaker:2014,andersen:2015}. Interactions can also be modulated by frequency tuning constructive or deconstructive interference between different coupling paths~\cite{srinivasan:2011}. Direct couplers are more compact, but qubits connected to the coupler are more sensitive to noise on the tuning degree-of-freedom; there is instrinsic protection from tuning noise for a tunable bus when we operate in the dispersive limit. Tunable couplers of both varieties have been realized in several experiments: between two tunable qubits~\cite{chen:2014,harrabi:2009}, between a qubit and resonator~\cite{srinivasan:2011,allman:2014} and between resonators ~\cite{wulschner:2015,zakka:2011,wang:2011}.  

In this work we realize a tunable bus between high-coherence fixed-frequency qubits with a relative detuning much larger than the anharmonicity. To turn on the interaction between the qubits, we modulate the tunable bus at the qubit difference frequency (as theoretically proposed in several Refs.~\cite{bertet:2006,niskanen:2006,kapit:2015}) which causes a parametric oscillation of the qubit-qubit exchange coupling and activates a resonant XX+YY interaction~\cite{majer:2007,dewes:2012}. The exchange interaction causes a two-qubit oscillation between states with one excitation $|10\rangle$ and $|01\rangle$, i.e., qubit 1 (Q1) in the excited state, qubit 2 (Q2) in the ground state and vice-versa. Applying this interaction for 183~ns we demonstrate a universal two-qubit gate --- the iSWAP gate --- with 0.982 average gate fidelity. Unlike drive-activated gates, the exchange interaction strength does not decrease when the qubit-qubit detuning is larger than the anharmonicity. In particular, for the detunings of the device in this work, $854~\rm{MHz}$, the leading gate for fixed-frequency qubits, cross-resonance~\cite{sheldon:2015b}, would not be viable. Although we demonstrate the gate between a single pair of qubits, in general, multiple qubits can be coupled to a single bus since the interaction is resonant in the detuning between specific qubit pairs. Therefore, the iSWAP gate is promising for implementing larger circuits where a range of qubit frequencies will be needed to avoid crosstalk and addressing errors. In addition, the tunable bus architecture enables analog quantum simulation schemes requiring controllable interactions. In particular, with this type of coupling ZZ and XX-YY two-photon interactions can also be activated by adjusting the modulation frequency. Moreover, the tunable bus can be used to more efficiently realize surface code implementations requiring iSWAP gates~\cite{ghosh:2015}.

Our paper is organized as follows. In \S~\ref{sect:tunable_bus_th} we discuss the theory of the tunable bus device and in \S~\ref{sect:tunable_bus_exp} introduce our two-qubit device. In \S~\ref{sect:osc} we show two-qubit iSWAP oscillations using our device and prepare and characterize a Bell state. In \S~\ref{sect:iswap} we present our universal two-qubit iSWAP gate and characterize the gate using randomized benchmarking and quantum process tomography. We conclude with a discussion in \S~\ref{sect:disc}.

\section{Tunable Bus \label{sect:tunable_bus}}

\subsection{Theory \label{sect:tunable_bus_th}}
\begin{figure}[ht!]
\includegraphics[width=0.48\textwidth]{\figfolder{1a}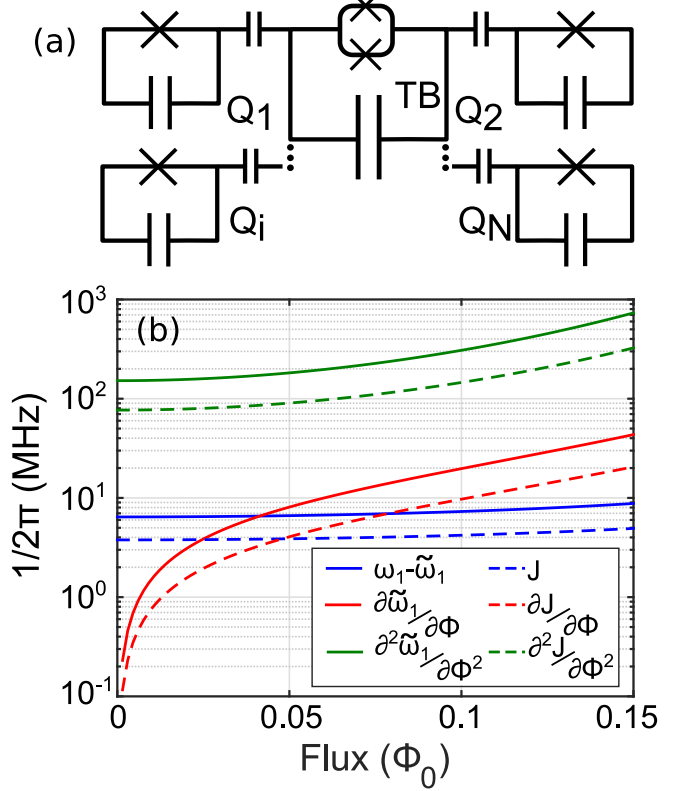}
\caption{(a) Schematic of an N qubit, 1 bus device. (b) Size of the expansion terms for $\omega$ and  $J$ versus the DC flux tuning the bus. Calculations are for the device parameters given in \S~\ref{sect:tunable_bus_exp}. \label{fig:1a}}
\end{figure}
The tunable bus circuit that we consider in this paper consists of several fixed-frequency qubits dispersively coupled to a frequency-tunable bus; a circuit schematic is shown in Fig.~\ref{fig:1a} (a). Because the bus is in the ground state and dispersively coupled, it suffices to keep only the first two levels of the bus. In terms of $N$ bare qubits coupled to a tunable bus, the standard circuit QED Hamiltonian is,
\begin{eqnarray}
	\frac{H}{\hbar} & = & \sum_{i=1}^{N} \left[ -\frac{\omega_{i}\hat{\sigma}_{i}^{Z}}{2} + g_i\left(\hat{\sigma}_{i}^{+}\hat{\sigma}_{TB}^{-} + \hat{\sigma}_{i}^{-}\hat{\sigma}_{TB}^{+}\right)\right] - \nonumber \\
	& & \frac{\omega_{TB}(\Phi) \hat{\sigma}_{TB}^{Z}}{2}, \label{eqn:bare}
\end{eqnarray}
where $\hat{\sigma}^{Z}$ is the Pauli-Z operator and $\hat{\sigma}^{+}$($\hat{\sigma}^{-}$) is the raising (lowering) operator. The tunable bus (TB) tunes with flux $\Phi$ as ~\cite{koch:2007},
\begin{equation}
	\omega_{TB}(\Phi) = \omega_{TB,0} \sqrt{\left|\cos(\pi \Phi /\Phi_0)\right|}, \label{eqn:tune}
\end{equation}
where $\Phi_0$ is the flux quantum. In the dispersive regime, i.e., $\left|g_i/(\omega_i-\omega_{TB})\right| \ll 1$, we can adiabatically eliminate the TB,
\begin{equation}
	\frac{H}{\hbar} = \sum_{i}^{N} -\frac{\tilde{\omega}_{i}(\Phi)\hat{\sigma}_{i}^{Z}}{2} + \sum_{j>i}^{N} J_{ij}(\Phi)\left(\hat{\sigma}_{i}^{+}\hat{\sigma}_{j}^{-} + \hat{\sigma}_{i}^{-}\hat{\sigma}_{j}^{+}\right), \label{eqn:dressed}
\end{equation}
thus realizing a flux tunable coupler. In Eq.~(\ref{eqn:dressed}) $\tilde{\omega}$ is the dressed qubit energy and $J_{ij}$ is the exchange coupling between qubits $i$ and $j$, which depend on flux as,
\begin{eqnarray}
	\tilde{\omega}_{i} & = & \omega_{i} + \frac{g_i^2}{\Delta_{i}(\Phi)}, \label{eqn:dressed_freq} \\
	J_{ij} & = & \frac{g_i g_j}{2} \left(\frac{1}{\Delta_{i}(\Phi)} + \frac{1}{\Delta_{j}(\Phi)}\right) \label{eqn:dressed_coup}
\end{eqnarray}
where $\Delta_{i}(\Phi) = \omega_{i}-\omega_{TB}(\Phi)$. To interact the qubits via the tunable coupler we apply a sinusoidal fast-flux bias modulation of amplitude $\delta$ so that the total flux applied to the tunable bus is $\Phi(t) = \Theta + \delta \cos(\omega_{\Phi} t)$. Expanding $\tilde{\omega}_i$ in the parameter $\delta \cos(\omega_{\Phi} t)$ to second-order where $\delta \ll 1$ we obtain
\begin{widetext}
\begin{eqnarray}
\tilde{\omega}_{i}(\Phi(t)) & \approx & \tilde{\omega_{\Phi}}_i(\Theta) + \left.\frac{\partial \tilde{\omega}_i}{\partial \Phi}\right|_{\Phi \rightarrow \Theta} \delta \cos(\omega_{\Phi} t) + \frac{1}{2}\left.\frac{\partial^2 \tilde{\omega}_i}{\partial \Phi^2}\right|_{\Phi \rightarrow \Theta} \left(\delta \cos(\omega_{\Phi} t) \right)^2,\\
& = &  \left[\tilde{\omega}_i(\Theta) - \frac{\delta^2}{4} \left.\frac{\partial^2 \tilde{\omega}_i}{\partial \Phi^2}\right|_{\Phi \rightarrow \Theta}\right] + \left.\frac{\partial \tilde{\omega}_i}{\partial \Phi}\right|_{\Phi \rightarrow \Theta} \delta \cos(\omega_{\Phi} t) + \frac{\delta^2}{4}\left.\frac{\partial^2 \tilde{\omega}_i}{\partial \Phi^2}\right|_{\Phi \rightarrow \Theta} \cos(2\omega_{\Phi} t).
\end{eqnarray}
Since the relation between qubit frequency and flux is nonlinear there is a second-order DC shift and an oscillating term at $2\omega_{\Phi}$; a similar expansion holds for $J_{ij}$. Typical values for these expansion terms are shown in Fig.~\ref{fig:1a} (b). In the frame rotating at the qubit frequencies for $\delta=0$ (the measurement frame), oscillating $\hat{\sigma}_Z$ terms and DC exchange coupling terms time-average to zero. Therefore, updating Eq.~(\ref{eqn:dressed}) to include all other expansion terms the Hamiltonian becomes
\begin{eqnarray}
	\frac{H}{\hbar} & = & \sum_{i}^{N} -\left(\tilde{\omega}_{i} - \frac{\delta^2}{4} \frac{\partial ^2 \tilde{\omega}_{i}}{\partial \Phi^2}   \right)\frac{\hat{\sigma}_{i}^{Z}}{2} +  \sum_{j>i}^{N}\left[\delta \frac{\partial J_{ij}}{\partial \Phi} \cos(\omega_{\Phi}t) - \frac{\delta^2}{4}\frac{\partial^2 J_{ij}}{\partial \Phi^2} \cos(2\omega_{\Phi} t)\right]\left(\hat{\sigma}_{i}^{+}\hat{\sigma}_{j}^{-} + \hat{\sigma}_{i}^{-}\hat{\sigma}_{j}^{+}\right), \label{eqn:dressed2}
\end{eqnarray}
where all values are evaluated at $\Phi=\Theta$. Because there is a drive-induced qubit shift, all N qubits will acquire a phase during the flux modulation pulse. This phase may be compensated after by applying single-qubit Z-gates. In a frame rotating at the qubit frequencies (including the drive-induced shift) the Hamiltonian is
\begin{eqnarray}
	\frac{H}{\hbar} & = &  \sum_{i,j>i}^{N}\left[\delta \frac{\partial J_{ij}}{\partial \Phi} \cos(\omega_{\Phi}t) + \frac{\delta^2}{4}\frac{\partial^2 J_{ij}}{\partial \Phi^2} \cos(2\omega_{\Phi} t)\right] e^{i \Delta_{ij,\delta} t}\left(\hat{\sigma}_{i}^{+}\hat{\sigma}_{j}^{-} + \hat{\sigma}_{i}^{-}\hat{\sigma}_{j}^{+}\right), \label{eqn:rotframe1}
\end{eqnarray}
\end{widetext}
where $\Delta_{ij,\delta} = \left(\tilde{\omega}_{i}-\tilde{\omega}_{j}\right) + \frac{\delta^2}{4} \left(\frac{\partial ^2 \tilde{\omega}_{j}}{\partial \Phi^2}-\frac{\partial ^2 \tilde{\omega}_{i}}{\partial \Phi^2} \right) $. When $\omega_{\Phi}$ is resonant with $\Delta_{ij,\delta}$ the Hamiltonian is
\begin{equation}
	\frac{H}{\hbar} = \frac{\delta}{2} \frac{\partial J_{ij}}{\partial \Phi} \left(\hat{\sigma}_{i}^{X}\hat{\sigma}_{j}^{X} + \hat{\sigma}_{i}^{Y}\hat{\sigma}_{j}^{Y}\right), \label{eqn:rotframe}
\end{equation}
which is a resonant exchange interaction between only qubits $i$ and $j$. There can also be a resonance condition when $2\omega_{\phi}=\Delta_{ij,\delta}$ with a different exchange coefficient. The interaction described by Eq.~(\ref{eqn:rotframe}) couples any states in the same excitation manifold. For two qubits this is only the set of states \{$|10\rangle,|01\rangle$\}. Applying this interaction for certain periods of time can generate entanglement and be used as a two-qubit gate. This will be explored in \S~\ref{sect:osc} and \S~\ref{sect:iswap}. 

\subsection{Experimental Device \label{sect:tunable_bus_exp}}

\begin{figure}[htp]
\includegraphics[width=0.45\textwidth]{\figfolder{1}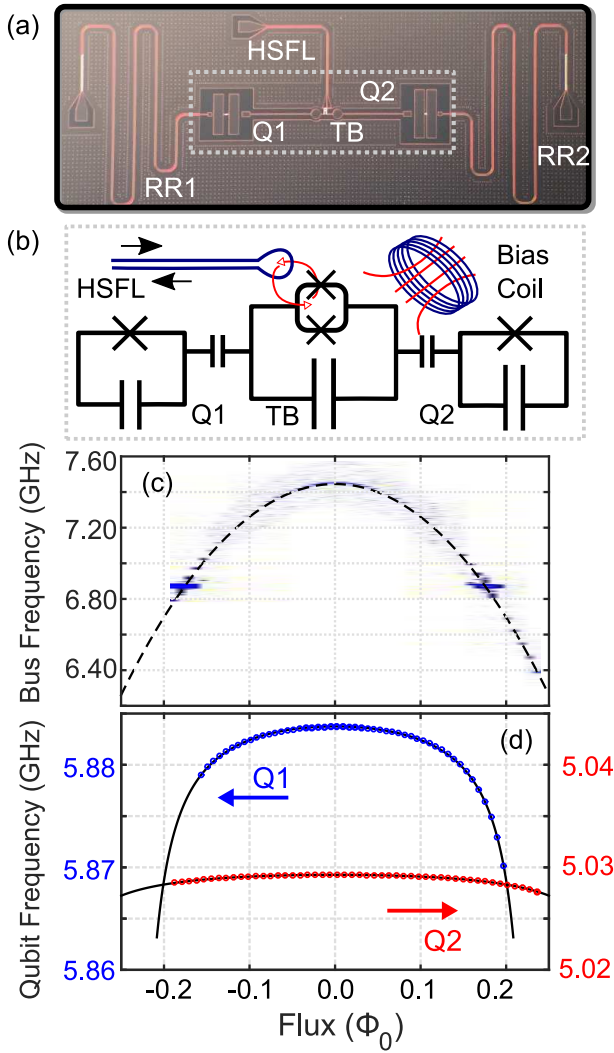}
\caption{(a) Optical image and (b) schematic of our circuit consisting of two fixed-frequency transmon qubits (Q1,Q2) coupled via a third tunable bus qubit (``tunable bus'' - TB). Q1 and Q2 have individual readout resonators (RR1,~RR2). The TB is tuned by a DC bias coil and a high-speed flux line (HSFL). Spectroscopy of (c) TB and (d) Q1,~Q2 frequency versus DC flux; Q1 tunes more strongly with flux because it is closer in frequency to the TB. We fit these tuning curves (solid lines) using Eq.~(\ref{eqn:dressed_freq}) to extract the Hamiltonian parameters for Eq.~(\ref{eqn:bare}). \label{fig:1}}
\end{figure}

Our experimental implementation of a two-qubit, one-bus device is shown in Fig.~\ref{fig:1}. By varying the DC flux, we can use Eq.~(\ref{eqn:tune}) and (\ref{eqn:dressed_freq}) to extract the bare Hamiltonian parameters $g_1(g_2)/2\pi=100.0(71.4)~\rm{MHz}$, $\omega_1(\omega_2)/2\pi = 5.8899(5.0311)~\rm{GHz}$, $\omega_{TB,0}/2\pi=7.445~\rm{GHz}$ by fitting to the measured frequencies  $\omega_{TB}$, $\tilde{\omega}_1$ and $\tilde{\omega}_2$ as shown in Fig.~\ref{fig:1}. These are transmon qubits with anharmonicity $\alpha/2\pi=-324(235)~\rm{MHz}$ where $\alpha$ is the detuning between the $|1\rangle \rightarrow |2\rangle$ transition and the qubit transition $|0\rangle \rightarrow |1\rangle$. One tradeoff of the tunable bus design is that the dressed qubits are susceptible to flux noise since the frequencies are flux-tunable. However, compared to a directly tunable qubit the flux noise sensitivity is lowered by a factor of $(g/\Delta)^2$. This noise has minimal effect on our 23~ns single-qubit gates. At the flux-bias point used to implement the two-qubit gate, $T_1=26.3(7) [50(3)]~\rm{\mu s}$ and $T_2=12.1(4)[28(1)]~\rm{\mu s}$ for Q1 [Q2] and the single qubit fidelities measured from randomized benchmarking are 0.99909(2) [0.99949(1)] (see Ref.~\cite{mckay:2015b} for benchmarking data and coherence measurements).

\section{Two-Qubit iSWAP Oscillations \label{sect:osc}}

\begin{figure}[htp]
\includegraphics[width=0.4\textwidth]{\figfolder{2}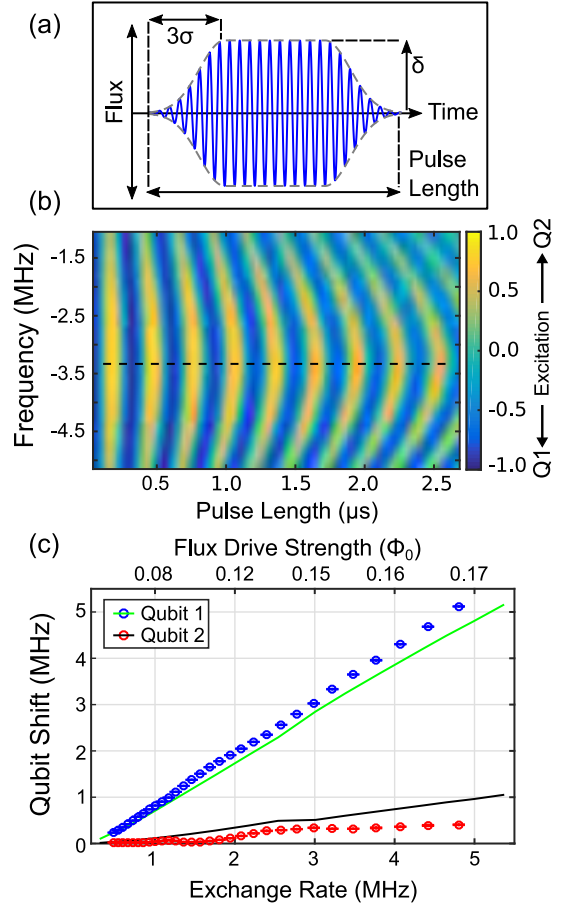}
\caption{(a) Flux modulation pulse of strength $\delta$. The pulse envelope (gray-dashed line) is a square pulse with Gaussian turn on and off with $\sigma=8.3$~ns and the turn on/off time is $3\sigma$. (b) Exchange oscillations between $|10\rangle$ and $|01\rangle$ as a function of the flux pulse length and the drive frequency $\omega_{\phi}$ (with respect to the detuning between the qubits when $\delta=0$, $\Delta_{12,\delta=0}\approx 854~\rm{MHz}$). This data is taken at a DC flux bias of $\Theta=-0.108\Phi_0$ with a constant flux pulse height $\delta=0.153\Phi_0$. The flux modulation induces a DC shift of the tunable bus and so the resonance frequency (dotted line) of the exchange oscillation is shifted down from $\Delta_{12,\delta=0}$ by approximately $3~\rm{MHz}$.  (c) Qubit shifts during the flux modulation pulse as a function of the drive strength (the procedure is described in the main text). The drive strength is plotted in terms of the measured exchange rate (bottom axis) and flux modulation strength $\delta$ in units of $\Phi_0$ (top axis) calculated theoretically from this exchange rate. The solid lines are no-free-parameters numerical calculations of the exchange rate and shift given a certain flux modulation by solving Eq.~(\ref{eqn:bare}). \label{fig:2}}
\end{figure}

To experimentally measure the exchange interaction we perform a $\pi$-pulse to prepare the state $|10\rangle$ (or $|01\rangle$) and then apply sinusoidal flux modulation pulses of strength $\delta$ and drive frequency $\omega_\phi$ in a range around $854~\rm{MHz}$ to couple the states and drive exchange oscillations. The flux pulse shape is shown in Fig.~\ref{fig:2} (a) and sample oscillations are illustrated in Fig.~\ref{fig:2} (b). In order to effectively drive these oscillations the tunable bus must be DC flux biased ($\Theta=-0.108\Phi_0$) since the strength of the exchange rate is proportional to the slope of the bus tuning curve, Eq.~(\ref{eqn:rotframe}). The slope is not purely linear so we also get a sizeable DC shift of the bus frequency, which in turn shifts the qubit frequency as given by Eq.~(\ref{eqn:dressed2}). We can measure the qubit shift during the oscillation by performing a Ramsey interferometric experiment: starting in $|00\rangle$, we apply a $\pi/2$ pulse to the qubit, then exchange for time $t$ (at a given flux modulation amplitude), reverse the exchange for time $t$ (flip the flux modulation pulse phase by 180 degrees) to return to the original state, then apply a final $\pi/2$ to the qubit. The fringe frequency measures the induced shift on the qubit frequency. The qubit shift versus the exchange rate is plotted in Fig.~\ref{fig:2} (c). Using the bare qubit parameters we construct a no-free-parameters theory curve (solid line). 

These qubit shifts have two important consequences for constructing a gate. For one, they are equivalent to applying single-qubit Z gates which therefore need to be compensated. Second, as we increase the exchange rate the coupler moves closer into resonance with the qubit. As discussed above, this reduces the protection to flux noise (i.e., $(g/\Delta)^2$ increases). Additionally, increasing the drive strength can lead to leakage out of the computational basis into the higher transmon levels and/or into the bus. Consequently, there is a tradeoff between coherence, leakage and exchange rate which puts an effective upper bound on how fast we can operate a two-qubit gate.  
 
\subsection{$\sqrt{\mathrm{iSWAP}}$ Bell State}

\begin{figure}[htp]
\includegraphics[width=0.4\textwidth]{\figfolder{3}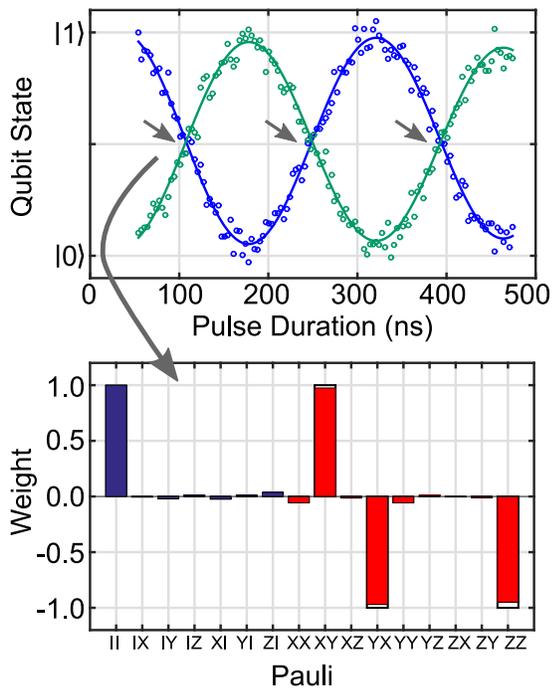}
\caption{(a) On-resonance SWAP oscillations between the qubits. We first apply a $\pi$-pulse to Q1 and then perform a variable length flux modulation pulse on the tunable bus with $\omega_{\phi}/2\pi=850.6~\mathrm{MHz}$ and $\delta=0.155\Phi_0$. The excitation oscillates between the two qubits and at special evolution times (indicated by arrows) entangled states are generated. (b) State tomography of the first entangled state in the Pauli representation with a fidelity of 0.974. Single qubit terms are illustrated in blue and two-qubit terms in red. The outlined bars show the ideal state $|\Psi\rangle=(|10\rangle-i|01\rangle)/\sqrt{2}$. \label{fig:3}}
\end{figure}

Specific flux modulation pulse lengths in Fig.~\ref{fig:2} represent primitive two-qubit gates that can be used to construct a universal gate set for quantum computing. As indicated by the arrows in Fig.~\ref{fig:3} (a) there are certain locations, $\Omega t= \pi(1/4+n/2)$ ($\Omega$ is the exchange rate), where the excitation is equally shared between both qubits. At these points a maximally entangled Bell state can be generated. At the first such crossing it is possible to realize a $\sqrt{i\mathrm{SWAP}}$ gate or iSWAP-$\pi/2$. Applying a $\sqrt{i\mathrm{SWAP}}$ gate to either the state $|10\rangle$ or $|01\rangle$ generates a maximally entangled Bell state. We perform state tomography on such a state and measure a fidelity of 0.974 as illustrated in Fig.~\ref{fig:3} (b).

\section{iSWAP Gate \label{sect:iswap}}
The $\sqrt{i\mathrm{SWAP}}$ gate is not in the Clifford group and so is not suitable for randomized benchmarking or as an error correction primitive. By extending the modulation pulse so that $\Omega t=\pi/2$ we realize the iSWAP gate (iSWAP-$\pi$), which is in the Clifford group. This gate swaps the states $|10\rangle,|01\rangle$ with a 90 degree phase with respect to the $|00\rangle,|11\rangle$ states which are unchanged (see inset to Fig.~\ref{fig:4}). The fidelity of this gate is sensitive to parasitic ZZ type interactions (e.g., a controlled phase). However, because the detuning between our qubits is large the ZZ interaction is only $66~\rm{kHz}$.  

\subsection{Gate Optimization and Simulation}

\begin{figure}[htb!]
\includegraphics[width=0.4\textwidth]{\figfolder{3a}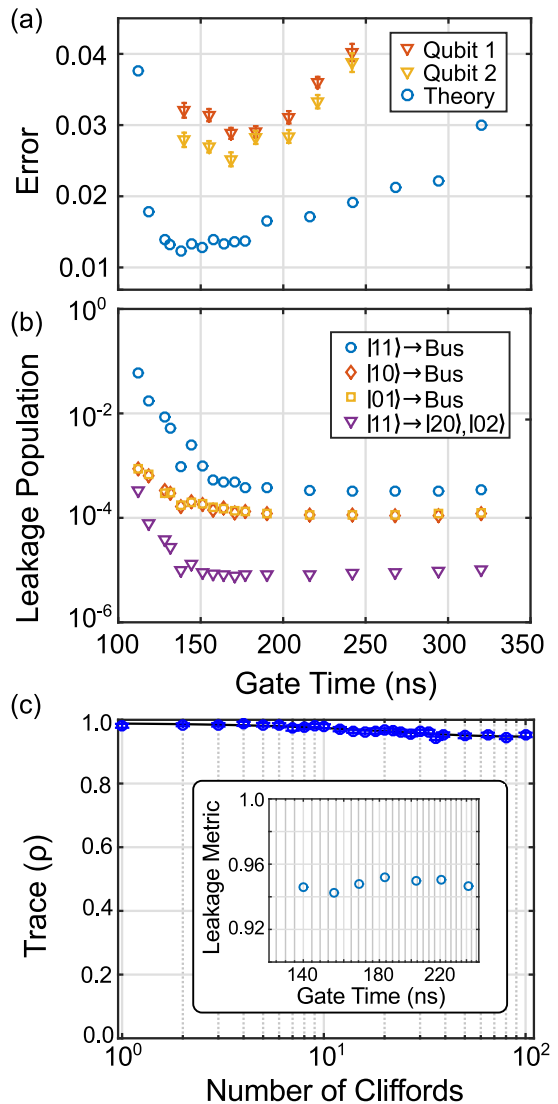}
\caption{Optimization and simulation of the iSWAP gate. (a) Gate error versus pulse width measured experimentally (triangles) versus the numerical calculation (circles) including levels outside the computational basis and decoherence. (b) Calculated leakage versus pulse width for different levels outside the computational basis. (c) Leakage measured experimentally by proxy by tracing over the computational states (leakage RB). This data is for the standard gate length 183~ns. (Inset) is the leakage metric (the asymptote of the RB data) versus pulse width. Because the leakage metric is flat versus pulse width we are not in the high leakage regime seen in the numerics (b). \label{fig:3a}}
\end{figure}

To first optimize the iSWAP gate we compare the gate error from two-qubit randomized benchmarking (RB) (see \S~\ref{sect:gatechar}) and simulation versus the gate length for $\Theta=-0.108\Phi_0$. For the numerics we model the system as Duffing oscillators (truncated to three levels in the calculation), as given by the Hamiltonian
\begin{eqnarray}
\label{HN}
H_N&=&\sum_{i=1}^2\left[\omega_i a^\dag_i a_i -\frac{\alpha_i}{2}(1-a^\dag_ia_i)a^\dag_i a_i\right]+ \nonumber \\
   & & \omega_{TB}(\Phi(t)) a^\dag_{TB} a_{TB} -\frac{\alpha_{TB}}{2}(1-a^\dag_{TB}a_{TB})a^\dag_{TB}a_{TB}\nonumber\\
&+&\sum_{i=1}^2g_{i}(a^\dag_i+a_i)(a^\dag_{TB}+a_{TB}),
\end{eqnarray}
which is the transmon generalization of Eq.~(\ref{eqn:bare}). Here we define creation (annihilation) operators for the $i$th fixed frequency qubit $a^\dag_i$ ($a_i$), with $0-1$ level transition energies $\omega_i$ and anharmonicities $\alpha_i$. Similar definitions are given for the tunable bus, with operators $a^\dag_{TB}$ ($a_{TB}$), and time-dependent frequency $\omega_{TB}(\Phi(t))$. The bus frequency as a function of flux is given by Eq.~(\ref{eqn:tune}) and the time-dependent flux pulse is the same shape as in the experiment as shown in Fig.~\ref{fig:2} (a). For the calculation we work in the measurement basis obtained by numerically diagonalizing Eq.~(\ref{HN}) when $\omega_{TB}(\Phi(t))=\omega_{TB}(\Theta)$. The unitary transformation to the measurement basis from $H_{N,0}$ is given by $U_{N,0}$. In a rotating frame at the dressed qubit frequencies the dynamics of the time-dependent flux pulse are described by the interaction Hamiltonian,
\begin{eqnarray}
	H_I(t) & = &  U_{I} \left[\omega_{TB}(\Phi(t))-\omega_{TB}(\Theta)\right] a^\dag_{TB} a_{TB} U_{I}^{\dagger}, \label{eqn:int} \\
	U_I & = &  e^{-i \left(U_{N,0}^{\dagger}H_{N,0} U_{N,0}\right) t} U_{N,0}.
\end{eqnarray}
For both the experiment and simulation we calibrate $\delta$ and $\omega_{\Phi}$ for a fixed pulse length. Experimentally, $\omega_{\phi}$ is calibrated by optimizing the oscillation contrast and $\delta$ by minimizing the error in the two-qubit rotation angle via error amplification techniques. The simulation parameters are calibrated numerically by evolving the state ${\ket{01}}$ by $H_I$ for a fixed gate time to state $|\Psi\rangle$ and optimizing the overlap $|\langle \Psi|10\rangle|^2$ (1 for a perfect iSWAP), as a function of the drive amplitude $\delta$ and drive frequency $\omega_{\Phi}$. The additional phases on the qubits in the measurement frame are also numerically and experimentally calibrated. \\
Using these procedures, we calibrate the gate experimentally and numerically for different gate times. Decoherence effects are included numerically by solving a master equation for the density matrix of the system
\begin{equation}
\label{MasterEq}
\dot\rho=-i[H_I,\rho]+\sum_{i=1}^2\left[\Gamma_{-,i}^{DC}\mathcal{D}[\sigma^-_i]\rho+\frac{\Gamma_{\phi,i}^{DC}}{2}\mathcal{D}[\sigma^Z_i]\rho\right].
\end{equation}
The superoperator $D[\hat{O}]\rho$ is defined in the standard way, $D[\hat{O}]\rho=(2\hat{O}\rho\hat{O}^{\dagger}-\hat{O}^{\dagger}\hat{O}\rho-\rho\hat{O}^{\dagger}\hat{O})/2$. The effective damping and Z operators $\sigma^-_i$, $\sigma^Z_i$ are defined in the measurement basis for the first two levels of the transmon qubits.  For each gate time we compute the average gate fidelity
\begin{equation}
F=\int d\Psi  \langle \Psi | U_{\textrm{iSWAP}}^{\dagger} \rho_{\ket{\Psi}} U_{\textrm{iSWAP}} |\Psi\rangle 
 \end{equation} 
 where $\rho_{\ket{\Psi}}$ is the resulting density matrix after evolving Eq.~(\ref{MasterEq}) with input state $\ket{\Psi}$ and $U_{\textrm{iSWAP}}$ is the ideal iSWAP gate.  There may be additional sources of error in the actual experiment such as 1/f flux noise and coupler losses, which are not considered in this calculation. Both the experimental and theoretical results for the gate error $1-F$ are shown in Fig.~\ref{fig:3a} (a). Numerically, we observe an optimal gate time of around $150$~ns. For shorter gate times the error rate increases substantially (likely due to leakage, which will be discussed next), while for longer times decoherence imposes a lower bound on gate error. Note that there are two sets of experimental measurements of gate error; one set is obtained by measuring the ground state of qubit 1 (tracing over qubit 2) and the other by measuring the ground state of qubit 2. It should be emphasized that these are from the same experiment, i.e. we perform a set of two-qubit Cliffords using the iSWAP gate as a primitive and then measure the average state of both qubit 1 and qubit 2 simultaneously through independent readouts. RB theory predicts that these measurements should give the same value for the fidelity since the random Clifford sequences mix errors equally to both qubits. However, we see a slight discrepancy between these two measurements that increases as we go to shorter gate times, e.g., at a gate length of 155~ns the error per gate differs by $4.4\times10^{-3}$. The source of said discrepancy is an ongoing investigation. Nevertheless, both measures of fidelity show the same trend and are consistent with the numerical data. The optimal fidelity for the experimental data suggests a slightly longer gate of approximately 180~ns and for further gate characterization (\S~\ref{sect:gatechar})  we select a gate time of 183~ns.\\

Increased error for short times is likely due to leakage out of the computational subspace. There are primarily two paths for leakage with this type of gate. The first path is a direct sideband drive from $Q1$ or $Q2$ to the tunable bus. This is a first-order process but is strongly off-resonance by ensuring that $|\Delta_{i,TB}|\gg|\Delta_{12}|$. The second path is from $|11\rangle \rightarrow |20\rangle,|20\rangle$ because our physical qubits are transmons and the resonant exchange interaction can also couple the set of states $\{|11\rangle,|20\rangle,|02\rangle\}$. The detuning of this transition compared to the wanted SWAP transition is,
\begin{eqnarray}
	|2\omega_{1/2}+\alpha_{1/2}-(\omega_1+\omega_2)|-|\Delta_{12}|, \\
	|\Delta_{12/21}+\alpha_{1/2}|-|\Delta_{12}|.
\end{eqnarray}
For large detuning compared to the anharmonicity this transition is off-resonant by the anharmonicity, which is large compared to the swap rate. For example in our sample $|\Delta_{12}|/2\pi=854~\mathrm{MHz}$ and $|\Delta_{12}+\alpha_1|/2\pi=530~\mathrm{MHz}$ and $|\Delta_{21}+\alpha_2|/2\pi=1089~\mathrm{MHz}$. From the numerics we can directly estimate leakage by evolving according to $H_I$ starting in the four basis states $\ket{00},\ket{01},\ket{10},\ket{11}$ and calculating the population in higher excited states after the gate. At short gate times, leakage is a considerable issue, but it becomes negligible as the gate time increases past $\approx$140~ns. The simulation results are shown in Fig.~\ref{fig:3a} (b).

 To characterize leakage experimentally we perform a variation of the RB process. First, we perform standard two-qubit RB and measure the average state of both qubits. The value measured on qubit 1 (normalized so that $|0\rangle$ is 1 and $|1\rangle$ is 0) is $\rho_{00}+\rho_{01}+\xi_{1}$ where $\xi_{1}$ represents leakage.  Next, we repeat the same experiment with a $\pi$-pulse at the end so that the measured state is now $\rho_{10}+\rho_{11}+\xi_{1}$ where $\rho$ is the density matrix just before the $\pi$-pulse and $\xi$ is unchanged by the pulse. Adding both qubits and measurements together we get,
\begin{eqnarray}
	& & (\rho_{00}+\rho_{01}+\xi_{1})+(\rho_{00}+\rho_{10}+\xi_{2})+ \nonumber \\
	& & (\rho_{10}+\rho_{11}+\xi_{1})+(\rho_{01}+\rho_{11}+\xi_{2}) \nonumber \\
	& = & 2 (\mathrm{Tr}(\rho) + \xi_{1} + \xi_{2}).
\end{eqnarray}
The exact values of $\xi_1,\xi_2$ are unknown because they depend on the leakage states, however, under the assumption they cause a deviation in the measurement signal we can look at this measure as a function of the RB sequence length to observe leakage trends. In Fig.~\ref{fig:3a} (c) we illustrate a representative leakage measurement for a 183~ns gate. Typical data asymptotes from one and we can define the asymptotic value to represent a leakage metric. Plotting the leakage metric versus gate length [inset of Fig.~\ref{fig:3a} (c)] we see that there is no strong evidence of leakage that is increasing as we decrease the gate length. We conclude that we are not in the strong leakage regime predicted by numerics and that leakage is not our limiting error. 

\subsection{Gate Characterization \label{sect:gatechar}}

\begin{figure}[ht!]
\includegraphics[width=0.4\textwidth]{\figfolder{4}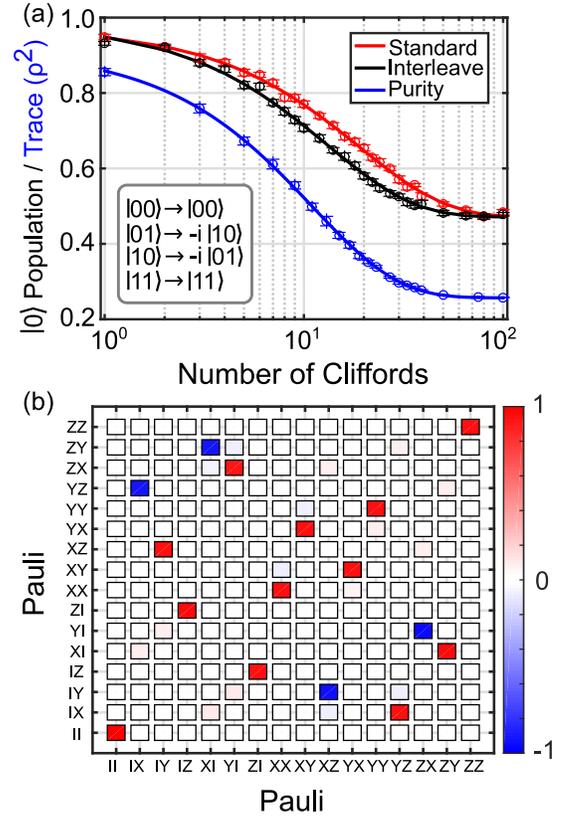}
\caption{(a) Standard, interleaved and purity randomized benchmarking (RB) of the two-qubit iSWAP gate with state transformation shown in the inset. For standard and interleaved RB the ground state population of qubit 2 is plotted as a function of the number of Clifford gates for a sample RB run consisting of 20 random seeds (14 seeds for the purity RB). The error numbers quoted in the main text are averaged over 8 such independent runs. For purity RB we plot the trace of $\rho^2$ versus the number of Clifford gates. (b) Pauli transfer matrix of the -iSWAP gate measured from quantum process tomography. \label{fig:4}}
\end{figure}

Finally, we perform full characterization of our optimal gate-length iSWAP of 183~ns with both randomized benchmarking (RB)~\cite{magesan:2011} and quantum process tomography. When composing two-qubit Clifford gates for RB from this gate set, on average there are 1.5 iSWAP gates per Clifford. The ground state population of Q2 as a function of the number of Cliffords is shown in Fig.~\ref{fig:4} (a). If we assume that the error per Clifford is predominantly due to the iSWAP gate, i.e. the single-qubit gates are effectively perfect, the error per gate averaged over 8 independent RB runs of 20 random seeds is $2.77(1)\times 10^{-2}$. A more direct error measurement is obtained by interleaved RB~\cite{magesan:2012}, as also shown in Fig.~\ref{fig:4} (a). Comparing the decay of the interleaved curve to the standard RB curve we extract an error of $1.77(4)\times 10^{-2}$ (with systematic error bounds of [0,0.08]). The measured error differs by approximately $2\times 10^{-3}$ whether we fit to the average ground state population of Q1 or Q2 (these are measured in the same experiment). Here we have quoted the more conservative of the two values. We also perform full quantum process tomography (QPT) on the gate as shown in Fig.~\ref{fig:4} (b). For this measurement we use 8000 measurements with a readout fidelity of 0.70(0.73) for Q1 (Q2). The fidelity from QPT is 0.949 from maximum likelihood estimation and 0.96 from the raw linear inversion. While QPT gives a full description of the gate in terms of the Pauli transfer matrix it is susceptible to state preparation and measurement (SPAM) errors. 

The estimated gate error from simulation is $1.5\times10^{-2}$, which is slightly lower than the measured error. To confirm that the discrepancy between the measured and calculated error rate is not due to coherent gate errors, e.g., calibration, we perform purity RB~\cite{wallman:2015} as shown in Fig.~\ref{fig:4} (a). For purity RB we measure the trace of $\rho^2$ ($\rho$ is measured from state tomography) after the RB sequence; these are the same sequences used for the standard RB measurement. Assuming pure depolarizing noise $\gamma$, the density matrix after $n$ Cliffords (starting in the ground state density matrix $\rho_0$) is,
\begin{eqnarray}
	\rho(n) & = & \gamma^n \rho_0 + (1-\gamma^n)\frac{\mathcal{I}}{d}, \\
	\rho^2(n) & = & \gamma^{2n} \rho_0^2 + (1-\gamma^n)^2 \frac{\mathcal{I}}{d^2} + \nonumber \\
			 & & 2\gamma^n(1-\gamma^n) \frac{\rho_0}{d}, \\
	\mathrm{Tr}(\rho^2[n]) & = & \gamma^{2n} + \frac{(1-\gamma^n)^2}{d} + \frac{2 \gamma^n (1-\gamma^n)}{d}, \\
			 & = & \left(1-\frac{1}{d}\right) \gamma^{2n}  + \frac{1}{d}. 
\end{eqnarray}
Therefore, we fit the data to $A\gamma^{2n}+B$ and label the quantity $\epsilon=\frac{3}{4}(1-\gamma^{\frac{2}{3}})$ as the purity error (per iSWAP gate). This procedure gives $\epsilon=2.2\times10^{-2}$, comparable to our gate error, demonstrating that our gate is dominated by incoherent errors.

\section{Discussion \label{sect:disc}}

In this paper we demonstrated a high-fidelity universal two-qubit gate by parametric modulation of a tunable bus. Importantly, the strength of the gate is not a strong function of the detuning between the qubits $\Delta_{ij}$. In contrast, drive-activated gates couple between manifolds, so invariably the higher-level states are coupled into the computational basis by the drive. As a result, the strength of the two-qubit terms decrease when $\Delta_{ij}$ is larger than $\alpha$ because, from the perspective of one qubit, the other qubit appears increasingly harmonic. For the device presented in this work $\Delta_{12,\delta=0}/2\pi=854~\rm{MHz}$ the leading drive-activated gate, cross-resonance, would not be viable~\cite{chow:2011}. As quantum circuits scale up it will be important to have qubits far apart in frequency to prevent addressibility errors and crosstalk. For example, calculations on the cross-resonance gate with several qubits coupled to the same bus indicate that there are number of unwanted resonant detuning conditions between pairs which will be difficult to avoid with qubits spaced closer than $\alpha$~\cite{magesan:2016,takita:2016}.

There is room for improvement in the gate error we measured. Since the error was effectively coherence-limited we could decrease the gate time or increase coherence. Decreasing the gate time may be difficult because of the leakage issues observed in simulation. Increasing the exchange coupling by increasing the qubit-bus coupling $g$ may also be difficult; this could also increase leakage and will certainly increase the parasitic ZZ interaction. Optimizing the gate time is an area for more consideration. Increasing coherence is less problematic and for coherences measured in comparable devices at IBM $T_1=T_2=80\mu s$~\cite{ibmqex}, gate errors should be  $<5\times10^{-3}$ and competitive with the best reported two-qubit gate errors $9\times10^{-3}$~\cite{sheldon:2015b} and $6\times10^{-3}$~\cite{barends:2014}. 

As discussed in \S~\ref{sect:tunable_bus_th} there is no fundamental limit to the number of qubits that can be coupled via a tunable bus since the coupling occurs resonantly at the detuning between pairs. Understanding the role of noise, crosstalk and operability with multiple qubits coupled to the same tunable bus is therefore an important open question for this architecture. In particular, four qubits coupled through a single tunable bus could serve as a surface code unit cell.

\begin{acknowledgments}
We acknowledge Firat Solgun, George Keefe and Markus Brink for simulation, layout and fabrication of devices. We acknowledge useful discussions with Sarah Sheldon, Lev Bishop, Jared Hertzberg and Matthias Steffen. This work was supported by ARO under contract W911NF-14-1-0124.
\end{acknowledgments}

\bibliography{swap_gate}

\end{document}